\documentclass[aps,prc,twocolumn,showpacs,superscriptaddress]{revtex4}
\usepackage{epsfig}
\usepackage{graphicx}
\usepackage{amsmath,amssymb,amsfonts}
\usepackage{array}
\usepackage{url}
\usepackage{hyperref}
\usepackage{multirow}
\usepackage{float}
\usepackage{lineno}
\usepackage{xspace}
\usepackage[usenames,dvipsnames]{color}
\newcommand{\sqsn}{\mbox{$\sqrt{s_{_{\rm NN}}}$}\xspace}
\newcommand{\bef}{\begin{figure}}
\newcommand{\eef}{\end{figure}}
\newcommand{\bc}{\begin{center}}
\newcommand{\ec}{\end{center}}
\newcommand{\nn}{\nonumber}
\newcommand{\be}{\begin{equation}}
\newcommand{\ee}{\end{equation}}
\newcommand{\bea}{\begin{eqnarray}}
\newcommand{\eea}{\end{eqnarray}}
\newcommand{\auau}{\mbox{Au$+$Au}\xspace}

\begin{document}
\title{Estimation of Stopped Protons at Energies Relevant for a  Beam Energy Scan\\ at the BNL Relativistic Heavy-Ion Collider}
\author{Dhananjaya Thakur}
\author{Sunil Jakhar}
\affiliation{Discipline of Physics, School of Basic Sciences, Indian Institute of Technology Indore, Indore- 453552, INDIA}
\author{Prakhar~Garg}
\affiliation{Discipline of Physics, School of Basic Sciences, Indian Institute of Technology Indore, Indore- 453552, INDIA}
\affiliation{Department of Physics and Astronomy, Stony Brook University, SUNY, Stony Brook, NY 11794-3800, USA}
\author{Raghunath Sahoo}
\email{Raghunath.Sahoo@cern.ch (Corresponding Author)}
\affiliation{Discipline of Physics, School of Basic Sciences, Indian Institute of Technology Indore, Indore- 453552, INDIA}
\begin{abstract}

The recent net-proton fluctuation results of  the STAR experiment from beam energy scan (BES) program at RHIC have drawn much attention to explore the QCD critical point  and the nature of deconfinement phase transition. There have been many speculations that the non-monotonic behaviour of $\kappa\sigma^{2}$ of the produced protons around $\sqrt{s_{\rm NN}}$ = 19.6 GeV in STAR results may be due to the existence of QCD critical point. However, the experimentally measured proton distributions contain protons from heavy resonance decays, from baryon stopping and from direct production processes. These proton distributions are used to estimate the net-proton number fluctuation. As it is difficult to disentangle the protons from the above mentioned sources, it is better to devise
a method which will account for the directly produced baryons (protons) to study the dynamical fluctuation at different center-of-mass energies. This is because, it is assumed that any associated criticality in the system could affect the particle production mechanism and 
hence, the dynamical fluctuation in various conserved numbers. In the present work, we demonstrate a method to estimate the number of stopped protons at RHIC BES energies for central ($0\%-5\%$) \auau collisions within STAR acceptance and discuss its implications on the net-proton fluctuation results.
 \end{abstract}
\date{\today}
\maketitle
\section{Introduction}
\label{intro}
One of the main motivations of  heavy-ion collisions is to explore the QCD phase diagram of the strong interaction. Quantum chromodynamics (QCD) predicts a phase transition from a hadron gas (HG) phase to a quark-gluon plasma (QGP) phase by varying the temperature (T) and/or baryon density ($\mu_{B}$) of the system. Lattice QCD calculations indicate a smooth crossover along the temperature axis, while various other models predict a first order phase transition at high
baryon density.  The existence of the QCD critical point is thus expected at finite $\mu_{B}$ and $T$, where the first order phase transition line ends  \cite{Stephanov:1998dy,Alford:1997zt,Stephanov:1996ki,Aoki:2006we,Fukushima:2010bq,Stephanov:2004wx,Fodor:2004nz, Stephanov:1999zu}. The search for the QCD critical point is one of the main motivations behind the recent STAR net-proton~\cite{Adamczyk:2013dal}, net-charge~\cite{Adamczyk:2014fia}
 and PHENIX net-charge~\cite{Adare:2015aqk} measurements. It is necessary to look into the dynamical behaviour of the produced system by considering the effects of baryon stopping, initial state participant fluctuations, and  their observable effects. In order to do this, the present work is an effort to quantify the effect of stopped baryons, which are prevalent at lower collision energies around the RHIC BES, where a possible critical point in the QCD phase diagram is expected to be observed.
\newline
  
 The phenomenon of baryon stopping could be used as a direct tool to explore the QCD phase transition and as a probe of the equation of state (EoS) of the system \cite{Ivanov:2010cu}. The reduced curvature obtained from the net-proton rapidity distribution at the mid-rapidity, is used as an observable for the baryon stopping. As discussed in Ref.\cite{Ivanov:2010cu}, the behaviour of the mid-rapidity curvature with $\sqrt{s_{\rm NN}}$ has been studied and a zig-zag type of structure is observed. A three-fluid dynamics (3FD) calculation with hadronic EoS fails to explain the observed structure, whereas 3FD with first order phase transition from a hadronic phase to a deconfined phase of quark-gluon plasma qualitatively reproduces the structure \cite{Ivanov:2010cu}. Hence, it is argued that the non-monotonic behaviour of the baryon stopping is due to a phase transition and is most probably of first order in nature.  Theoretical studies based on STAR net-proton fluctuation \cite{Adamczyk:2013dal} hint for a phase transition at low $\sqrt{s_{\rm NN}}$ taking inclusive protons, {\it i.e.} from both production and stopping. However, as baryon stopping plays a major role at lower collision energies and almost vanishes at higher energies, there seems to be a need to disentangle the contribution of stopped baryons and the produced baryons in order to understand the observed structure and hence, the QCD phase diagram.
 
The conserved number fluctuations are associated with the possible existence of critical point, which are dynamical in nature. The present work
is motivated to study net-baryon (proton) number fluctuation, which is related to particle production mechanism.
But, most of the experimentally measured proton distributions contain the protons from stopping and resonance decays besides from the direct production. Also, it has been studied earlier that the change in the mean of proton distributions (which will be there, after subtracting the stopped protons) will have large effect on the correlation of the protons and antiprotons, and it can influence the higher moments of net-protons fluctuations~\cite{Mishra:2015ueh}. Particularly, at lower center-of-mass energies the stopping contribution is most dominant and it will be interesting to look for the dynamical fluctuations after removing these stopped protons. Various other effects on the conserved number fluctuations have been studied earlier in Refs.~\cite{Mishra:2016qyj,Mishra:2015ueh, Garg:2013ata,Garg:2012nf,Garg:2015owa,Mishra:2013qoa}.

In heavy-ion collision experiments, a part of the incident energy of the two colliding nuclei is used for fireball production and hence, the production of secondary particles. Therefore, the formation of a QGP in relativistic heavy-ion collisions depends on the amount of stopping between the colliding ions particularly at low center-of-mass energies. Hence, baryon stopping serves as an important tool to understand the particle production mechanism. As the net-baryon number is conserved and rapidity distribution is modified due to re-scattering of the particles after the collision, the net-baryon rapidity distribution becomes a useful probe to give information about the baryon transport and baryon stopping. Since, the neutrons are not measured mostly in heavy-ion experiments, the net-proton rapidity distributions are used to quantify the baryon stopping \cite{Feng:2011zze, Bearden:2003hx, Videbaek:2005qn, Zhou:2009yt, Li:2011tt, Arsene:2009aa} .

In the present work, we use the data of net-proton rapidity distributions for the most central \auau collisions measured at AGS, SPS and by the BRAHMS experiment at RHIC. We use a two source function~\cite{Ivanov:2010cu} to analyze the net-proton distributions at different center-of-mass energies. It is a combination of two thermal sources with a shift in their rapidities. The same two source function is used in the present work to calculate the percentage of stopped protons. Afterwords, the percentage of stopped protons with $\sqrt{s_{\rm NN}}$ is parametrized and the values at RHIC  BES energies are interpolated. Finally, we estimate the contribution of stopped protons in  rapidity, $|y| ~< ~0.5$ and transverse momentum between 0.4 GeV/c  and 0.8 GeV/c, which are used to measure the protons and antiprotons by STAR experiment for net-proton fluctuation studies~\cite{Adamczyk:2013dal}.

The paper is organized as follows. In Section~\ref{sec:method}, we discuss the method used for the present analysis. It is divided into three sub-sections: (a) \hyperref[sec1] Estimation of baryon stopping in $|y| <  $ 0.5, (b) \hyperref[sec2] Estimation of stopped protons in STAR transverse momentum ($p_{T}$) range, and (c) \hyperref[sec3] Contribution of stopped protons in STAR measurements. We briefly discuss the implications of this work to the net-proton fluctuation results of STAR experiment at RHIC in Section~\ref{sec:expt}.  Finally in Section~\ref{sec:sum}, we summarize our work.

 \bef[ht!]
 \bc
 \includegraphics[scale=0.45]{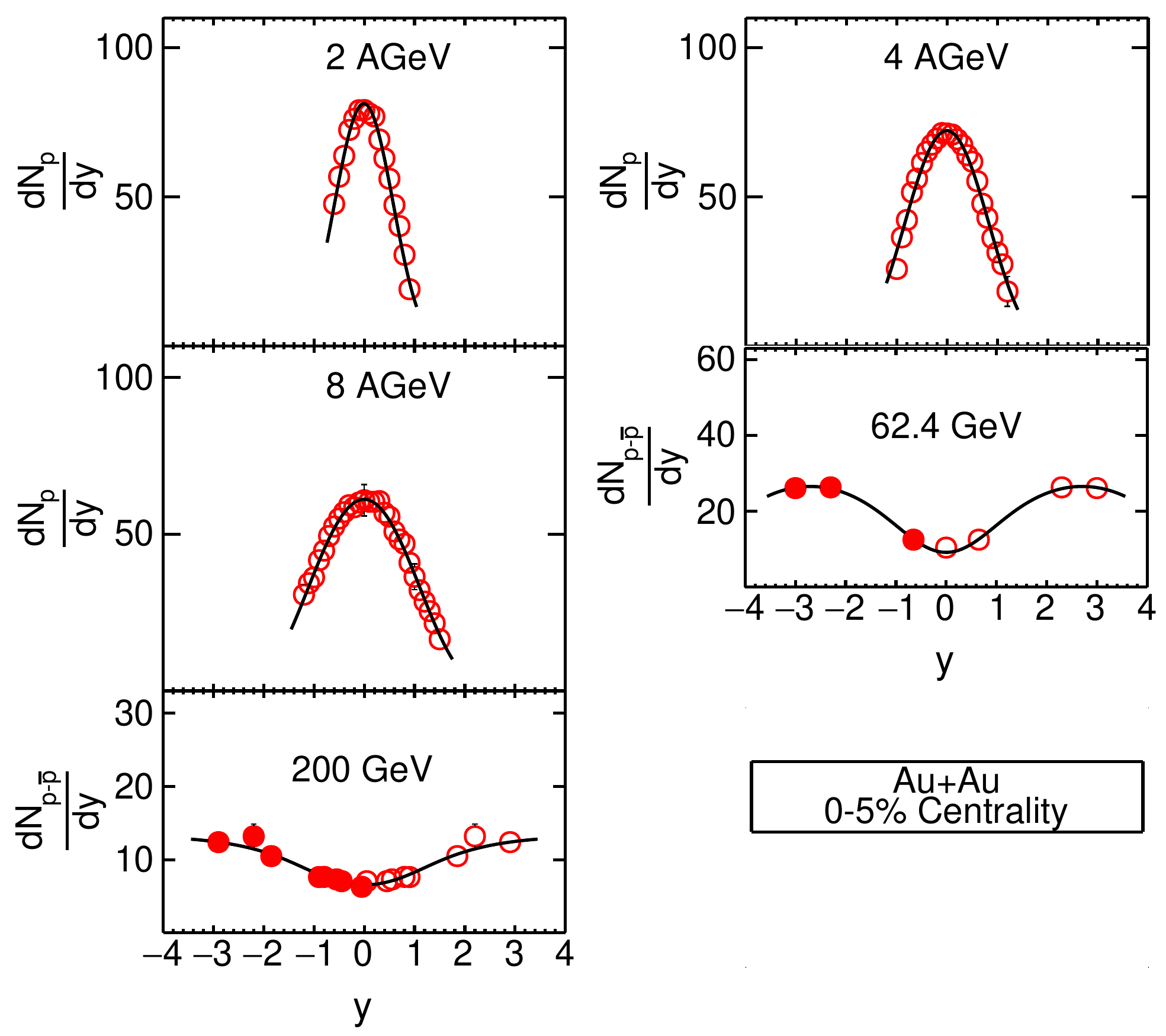}
 \caption  {The rapidity densities of protons at 2, 4, 8 AGeV (AGS) and net-proton ($N_{p}-N_{\bar{p}}$) (for RHIC energies) for central Au+Au collisions in the center-of-mass system. Experimental data are 
from E802 ~\cite{Ahle:1999in}, E877 ~\cite{Barrette:1999ry}, E917 ~\cite{Back:2000ru}, E866~\cite{Stachel:1998rc} and RHIC experiments \cite{Bearden:2003hx,Arsene:2009aa}. The open circles are experimentally measured data points and the filled circles are the mirror reflections, assuming a symmetry in particle production. Solid lines represent the two source fit function given by Eq.\ref{eq1}.}
 \label{fig1}  
 \ec
 \eef

\section{Method}
\label{sec:method}

\subsection{Estimation of baryon stopping in $|y| <  $ 0.5}
\label{sec1}
Net-proton rapidity distribution of the experimental data is best described by the following function:   
\bea
\label{eq1}
dN/dy = a~[exp\{-(1/w_{s}) \cosh(y-y_{cm} - y_{s})\} \nn\\  + exp\{-(1/w_{s}) \cosh(y-y_{cm}+y_{s})\}],
\eea

where a, $y_{s}$ and $w_{s}$ are the fit parameters of the function and $y_{cm}$ is the center-of-mass rapidity of the colliding nuclei~\cite{Ivanov:2010cu}. Eq.\ref{eq1} is the sum of two thermal sources shifted by rapidity $\pm y_{s}$ from the mid-rapidity.  $ w_{s}$ is the width of  the sources and is given by  $ w_{s}$ = (temperature)/(transverse mass), with the assumption that there is no spread of collective 
velocities in the sources with respect to the source rapidities.  We have used Eq.~\ref{eq1} to fit the net-proton rapidity distributions to quantify the baryon stopping. Since we are concerned about symmetric collisions, the parameters of the two sources are taken identical. Parameters $y_{s}$ and $w_{s}$ are calculated from the fitted function to the rapidity distribution of secondary particles. 

Baryon stopping is directly measured via the rapidity distribution of net-protons ({\it i.e.} the number of protons minus antiprotons). At low energies like AGS, the production of antiprotons is very small so the net-proton distribution is assumed to be the same as the proton distribution and the rapidity distribution peaks at mid-rapidity \cite{Ivanov:2010cu}. As the collision energy increases a dip begins to appear at mid-rapidity ($y \approx 0$) and  the peak shifts towards the  forward and backward rapidities, due to the production of antiprotons at mid-rapidity. This indicates that with increase of energy the transparency at the mid-rapidity increases. The rapidity loss of the particle is defined as $y_{loss} = y_{b} - y$, where $y_{b} = \ln(\sqrt{s_{\rm NN}}/m_{p})$ is the beam rapidity with $m_{p}$ being the mass of proton.

Therefore, the net-baryon number at mid-rapidity is the measure of baryon stopping~\cite{Feng:2011zze, Bearden:2003hx, Videbaek:2005qn}. To quantify the baryon stopping, we have used  Eq.~\ref{eq1} to fit the data of rapidity distribution for the most central Au+Au collisions at AGS (2, 4, 8 AGeV), and BRAHMS data at RHIC (62.4 and 200 GeV). The fitting is performed using TMinute class available in ROOT library with $\chi^{2}$ minimization. This is shown in FIG.~\ref{fig1} along with the fit functions for different energies. Further, these fit functions are used to estimate the baryon stopping at corresponding collision energies.

In the center-of-mass system, the maximum rapidity that an outgoing particle can have after the collision is $y_{b}$, which is only possible for full transparency. Therefore, by using Eq.~\ref{eq1}, the fraction of stopped protons in $|y| <  $ 0.5 at a particular energy can be calculated as,
 
\bea
\label{eq2}
f^{protons}_{stopped}=\frac{\int\limits_{-0.5}^{0.5} \frac{dN}{dy}dy}{\int\limits_{-y_{b}}^{y_{b}} \frac{dN}{dy}dy}\,.
\eea

Hence, the percentage of stopped protons can be estimated as:
\bea
\label{eq3}
N^{protons}_{stopped} = f^{protons}_{stopped}\times100 ~\%.
\eea

It should be noted here that  $\int\limits_{-y_{b}}^{y_{b}} \frac{dN}{dy}\,dy$ gives the total number of participating protons ($N^{B}_{part}$) and for the top central \auau collisions, $N^{B}_{part}\approx$ 158. The calculated percentage of stopped protons at these energies are shown in TABLE.~\ref{table1}.

  \begin{table}[ht]
   \centering
     \caption[]{Fraction of stopped protons in $|y| <  $ 0.5 at different energies}
     \begin{tabular}{|c|c|c|c|c|c|c|c|}
       \hline
   \sqsn [GeV]& $\rm E_{\rm lab} [GeV/nucleon]$& $\rm N^{\rm protons}_{\rm stopped}[\%]$              \\ \hline
   2.35 & 2       &~64.22 $\pm$ 0.10~          \\ \hline  
   3.04 & 4       &~52.16 $\pm$ 0.15~           \\ \hline
   4.09 & 8       &~44.61$\pm$ 0.12~          \\ \hline
   62.4  &    &~4.57$\pm$ 0.10~           \\ \hline
   200    &  &~2.91$\pm$ 0.0~                 \\ \hline
 
   \hline
     \end{tabular}
   \label{table1}
   \end{table}
 
\begin{figure}[ht!]
\centering
\includegraphics[scale=0.4]{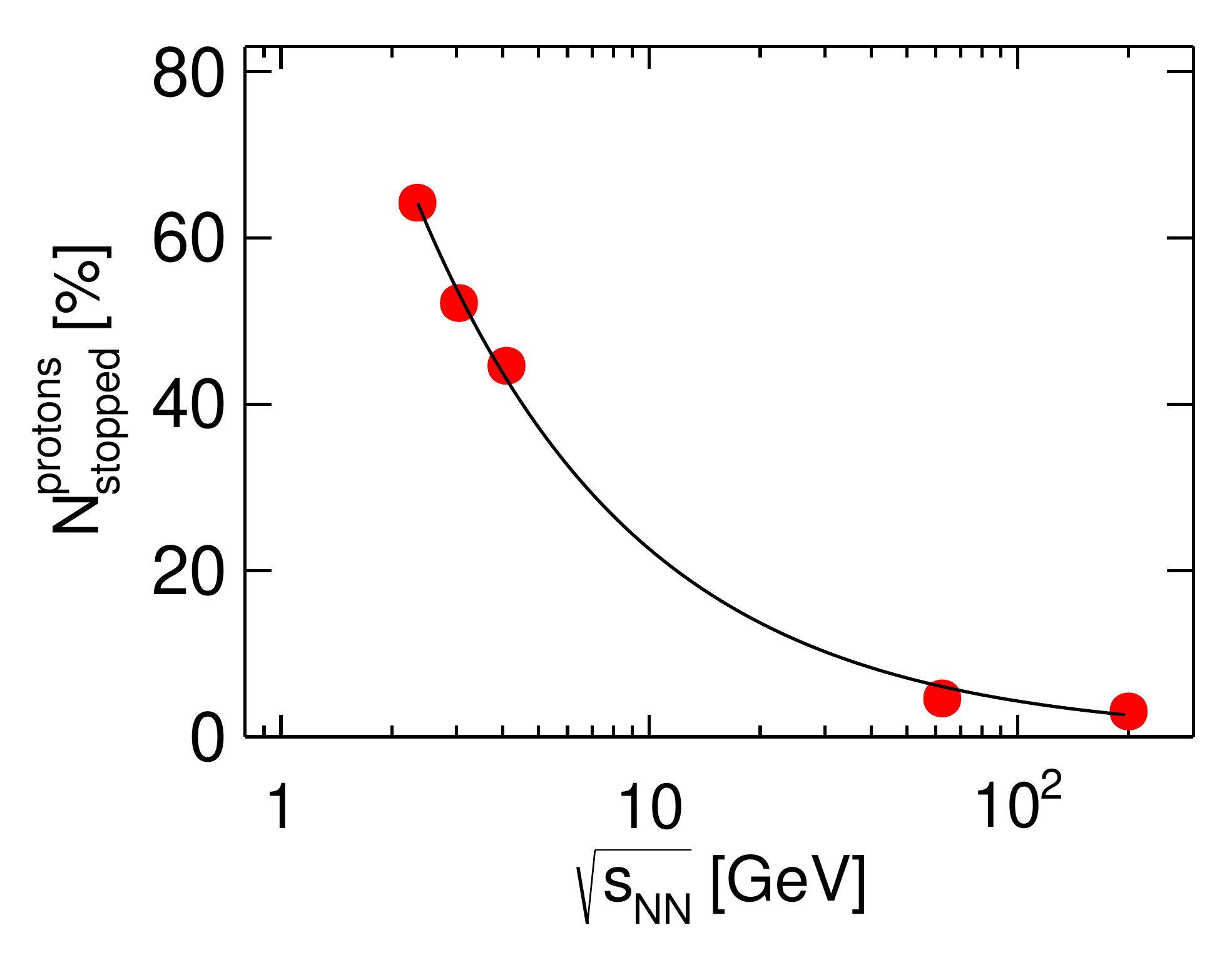}
\caption{Percentage of baryon stopping as a function of $\sqrt{s_{\rm NN}}$, fitted with an exponentially decreasing function with energy.}
\label{fig2}
\end{figure}

Thereafter, by converting the lab energy to the center-of-mass energy, a study of stopped protons with $\sqrt{s_{\rm NN}}$ is done. The decreasing behaviour of stopped protons ($N^{\rm proton}_{\rm stopped}$) with $\sqrt{s_{\rm NN}}$ is best described by an exponential function as shown in FIG.~\ref{fig2}. Our observations go inline with the earlier study~\cite{Feng:2011zze}. Hence, a parametric form of the stopped protons with $\sqrt{s_{\rm NN}}$ is obtained. This is given by, $\rm N^{\rm protons}_{\rm stopped}[\%]$ $= A \exp(-B\ln(\sqrt{s_{\rm NN}/s_0}))$, where $\sqrt{s_0}$ is taken to be 1 GeV and the obtained fitting parameters are $A= 118.89 \pm 6.18$ and $B= 0.72\pm 0.05$. Using this parametrized function, we have interpolated the percentage of stopping at RHIC BES energies in $|y|<0.5$ for \sqsn= 7.7, 11.5, 19.6, 27, 39, 62.4 and 200 GeV. The values are tabulated in TABLE~\ref{table2}.

 \begin{table}[ht]
\centering
 \caption{Percentage of stopped protons at different RHIC BES energies in $|y| <  $ 0.5 acceptance.}
    \begin{tabular}{|c|c|c|c|c|c|c|c|}
\hline
$\sqrt{s_{\rm NN}}$ [~GeV] & 7.7  & 11.5  & 19.6  & 27 & 39 & 62.4 & 200\\ \hline
$\rm N^{\rm protons}_{\rm stopped}[\%]$& 27.27  &20.43 &13.90    &11.03   &8.46  &6.03   &2.60 \\ \hline 
$\rm ~Err\{N^{protons}_{stopped}[\%]\}$& 1.30  &1.34 &1.25    &1.15   &1.03 &0.86   &0.51 \\ \hline 
   \end{tabular}
  \label{table2}
  \end{table}

\subsection{Estimation of stopped protons in STAR $p_{T}$ acceptance} 
\label{sec2}
 In the previous section, we have estimated percentage of stopped protons in $|y| <  $ 0.5 at RHIC BES energies. But, the STAR experiment has measured the protons and antiprotons in 0.4 GeV/c  $< p_{T} <$  0.8 GeV/c to calculate the higher order cumulants of net-proton distributions~\cite{Adamczyk:2013dal}. After estimating the number of stopped protons at mid-rapidity, now we calculate the same in the STAR $p_{T}$-acceptance.  Here we assume that the stopped protons are uniformly distributed over the whole $p_{T}$-spectra. To estimate the fraction of the stopped protons in STAR $p_{T}$-acceptance we have fitted the protons $p_{T}$-spectra at different available BES energies. The fitting is performed with a Levy-Tsallis function, which is given by Eq.11   of Ref. \cite{Abelev:2006cs}  for $\sqrt{s_{\rm NN}}$ = 9.2, 62.4 and 200 GeV. Similarly, for $\sqrt{s_{\rm NN}}$ = 19.6 GeV, the Tsallis distribution function, as given by Eq.6 of Ref. \cite{Thakur:2016boy} is used. The fitting results of STAR proton data are shown in Figs.~\ref{fig3a} and \ref{fig3b}. The ratios in the lower panels of Figs.~\ref{fig3a} and \ref{fig3b} suggest that the used fit functions describe the experimentally measured proton spectra very well for all centre-of-mass energies. We integrate the fitted functions at a particular energy from $p_{T}$ = 0.0 to $p_{T}=\infty$ and $p_{T}$ = 0.4 to 0.8 (GeV/c). Thus the total number of protons in the whole $p_{T}$ range and the number of protons in 0.4 GeV/c  $< p_{T} <$  0.8 GeV/c are calculated.
 
 \bef[H]
 \bc
 \includegraphics[scale=0.54]{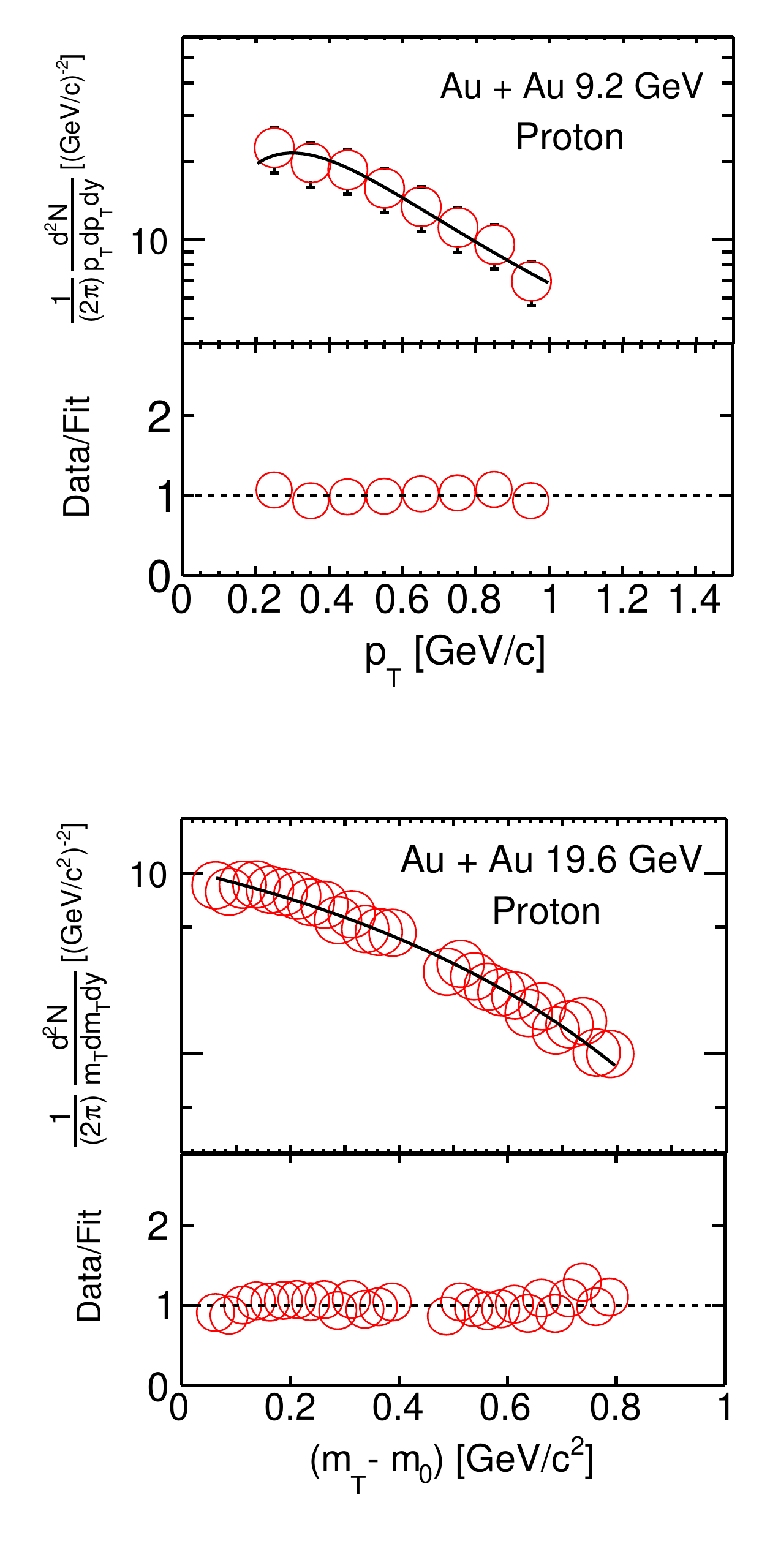}  
 \caption{Invariant yield of protons at $\sqrt{s_{\rm NN}}$ = 9.2 GeV \cite{Abelev:2009bw}, 19.6 GeV \cite{Picha:2005pq} for the most central Au + Au collisions. The open circles are the experimental data measured by the STAR Collaboration. The solid lines are the Levy-Tsallis function \cite{Abelev:2006cs}  for $\sqrt{s_{\rm NN}}$ = 9.2 and the Tsallis function \cite{Thakur:2016boy} for $\sqrt{s_{\rm NN}}$ = 19.6 GeV. The lower panels show the ratio of data points to their functional values.}
 \label{fig3a} 
 \ec
 \eef
 
 \bef[H]
 \bc
 \includegraphics[scale=0.45]{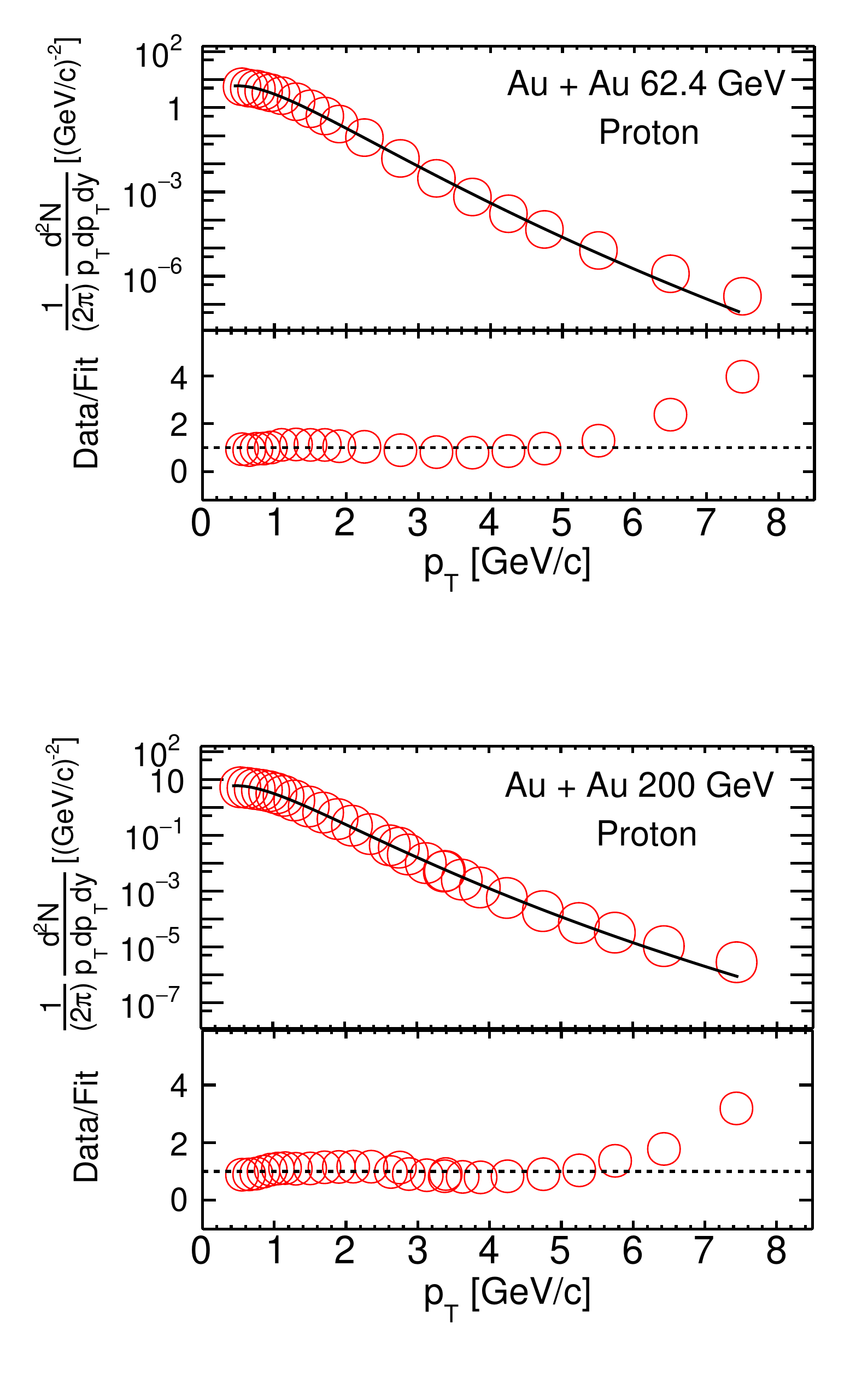}  
 \caption{Invariant yield of protons at 62.4 GeV \cite{Abelev:2007ra} and 200 GeV \cite{Abelev:2006jr} for the most central Au + Au collisions. The open circles are the experimental data measured by the STAR Collaboration. The solid lines are the Levy-Tsallis function \cite{Abelev:2006cs}. The lower panels show the ratio of data points to their functional values.}
 \label{fig3b} 
 \ec
 \eef

 The fraction of protons falling in 0.4 GeV/c  $< p_{T} <$  0.8 GeV/c region is estimated as,
 
  \begin{equation}
 f^{protons}_{p_{T}} = \frac{N^{protons}  (0.4<p_{T}<0.8)  } {N^{protons}(full~p_{T})}
 \end{equation}
 Hence, the percentage of protons in STAR $p_{T}$-acceptance is given by 
 \begin{equation}
N^{protons}_{p_{T}} = f_{p_{T}}^{protons}\times 100 \%
\label{frac}
  \end{equation}

Using Eqn. \ref{frac}, one can estimate the fraction of stopped protons contributing in STAR $p_{T}$-acceptance.
To calculate the percentage in the RHIC BES energies, we have parametrized these numbers with $\sqrt{s_{\rm NN}}$ by first order polynomial as shown in FIG.~\ref{fig4}. Then the contributions at different BES energies are interpolated. The extracted values are tabulated in TABLE~\ref{table3}. It is clear from the table that the fraction of stopped protons decrease with collision energy. 
 
  
 \bef[H]
 \bc
 \includegraphics[scale=0.40]{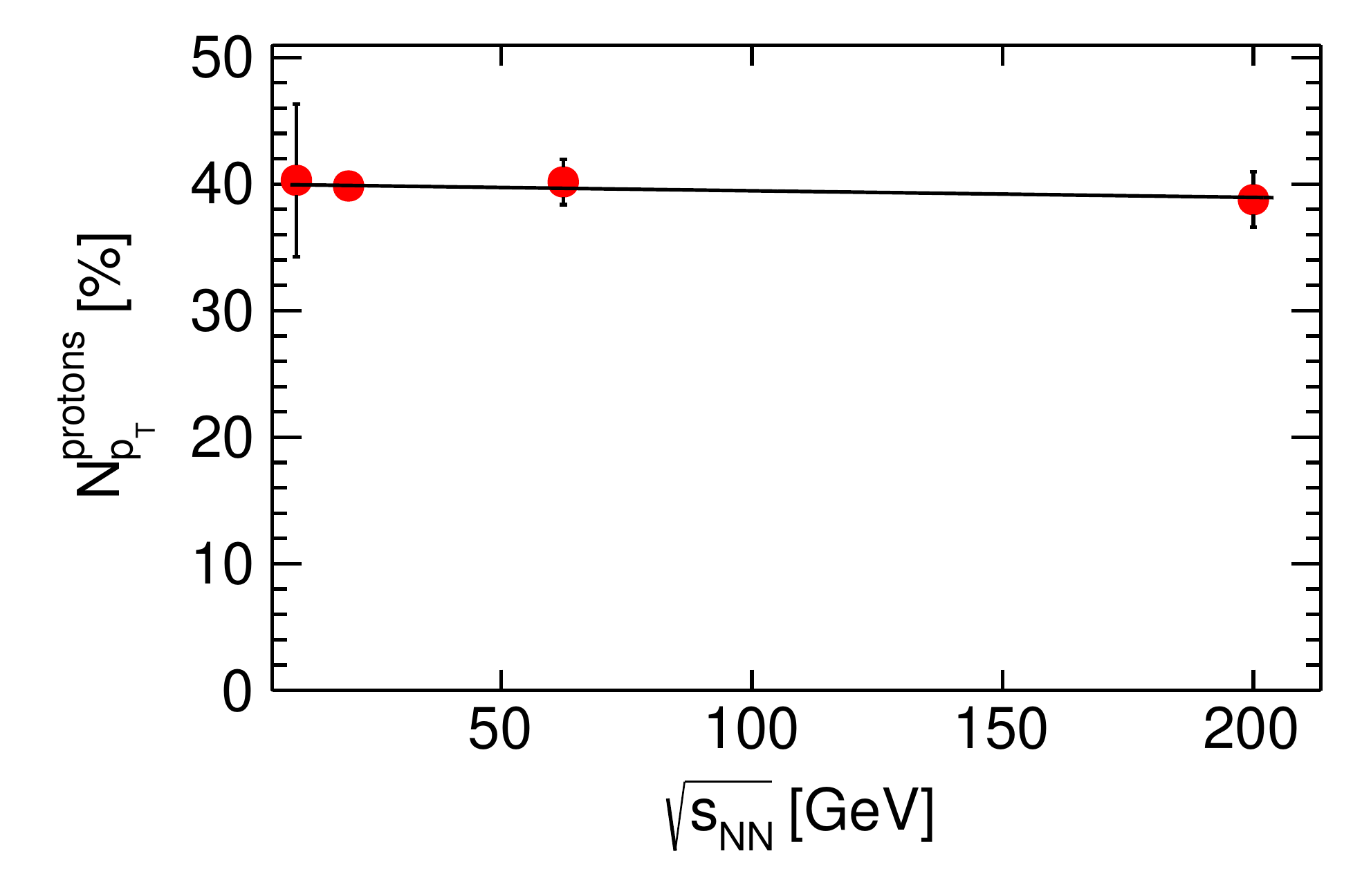}  
 \caption{Percentage of protons in 0.4 GeV/c  $< p_{T} <$  0.8 GeV/c as a function of $\sqrt{s_{\rm NN}}$ are being well described by a first order polynomial. }
 \label{fig4} 
 \ec
 \eef
  
\begin{table}[ht]
\begin{center}
\centering
 \caption{Percentage of protons obtained from $p_{T}$-spectra at different BES energies in STAR $p_{T}$-acceptance.}
\begin{tabular}{|c|c|c|c|c|c|c|c|}
\hline
$\sqrt{s_{\rm NN}}$ [GeV]& 7.7  & 11.5  & 19.6  & 27 & 39 & 62.4 & 200\\ \hline
$\rm N^{\rm protons}_{p_{\rm T}}\%$ &39.95  &39.93 &39.88   &39.84     &39.79  &39.66   &38.94  \\ \hline 
$\rm Err~\{N^{\rm protons}_{p_{\rm T}}\%\}$ &0.60  &0.58 &0.54   &0.52     &0.52  &0.63   &2.11  \\ 
 \hline
    \end{tabular}
    \label{table3}
 \end{center}
  \end{table}
 
  \subsection{Contribution of stopped protons in STAR measurements}
  \label{sec3}
  In the top central \auau collisions, there are around 158 protons participating in each collision. We have already estimated the number of protons in TABLE.~\ref{table1} in $|y| <  $ 0.5 for full $p_{T}$-coverage and in TABLE.~\ref{table3} for 0.4 GeV/c  $< p_{T} <$  0.8 GeV/c. Therefore, one can easily calculate the effect of both to estimate the total stopped protons in STAR acceptance. This is given by

\begin{equation}
  N^{protons}_{stopped}(STAR) = 158~\times~N^{protons}_{stopped}\%~\times~N^{protons}_{p_{T}}\%,
    \end{equation}
 where  $N^{\rm protons}_{\rm stopped}(\rm STAR)$ is the total contribution of stopped protons in STAR acceptance. In TABLE.~\ref{table4} we enlist the total number of stopped protons at BES energies in STAR acceptance.
 
 \begin{table}[ht]
\centering
 \caption{Number of stopped protons at different RHIC BES energies in STAR acceptance.}
    \begin{tabular}{|c|c|c|c|c|c|c|c|}
\hline
$\sqrt{s_{\rm NN}}$ [~GeV] & 7.7  & 11.5  & 19.6  & 27 & 39 & 62.4 & 200\\ \hline
$\rm N^{\rm protons}_{\rm stopped}(\rm STAR)$& 17.21  &12.89 &8.76   &6.94   &5.32  &3.78   &1.6 \\ \hline 
$\rm Err~\{N^{\rm protons}_{\rm stopped}(STAR)\}$& 0.86  &0.86 &0.80    &0.73  &0.65 &0.54  &0.33 \\ \hline 
   \end{tabular}
  \label{table4}
  \end{table}
  
\begin{figure}[ht!]
\centering
\includegraphics[scale=0.4]{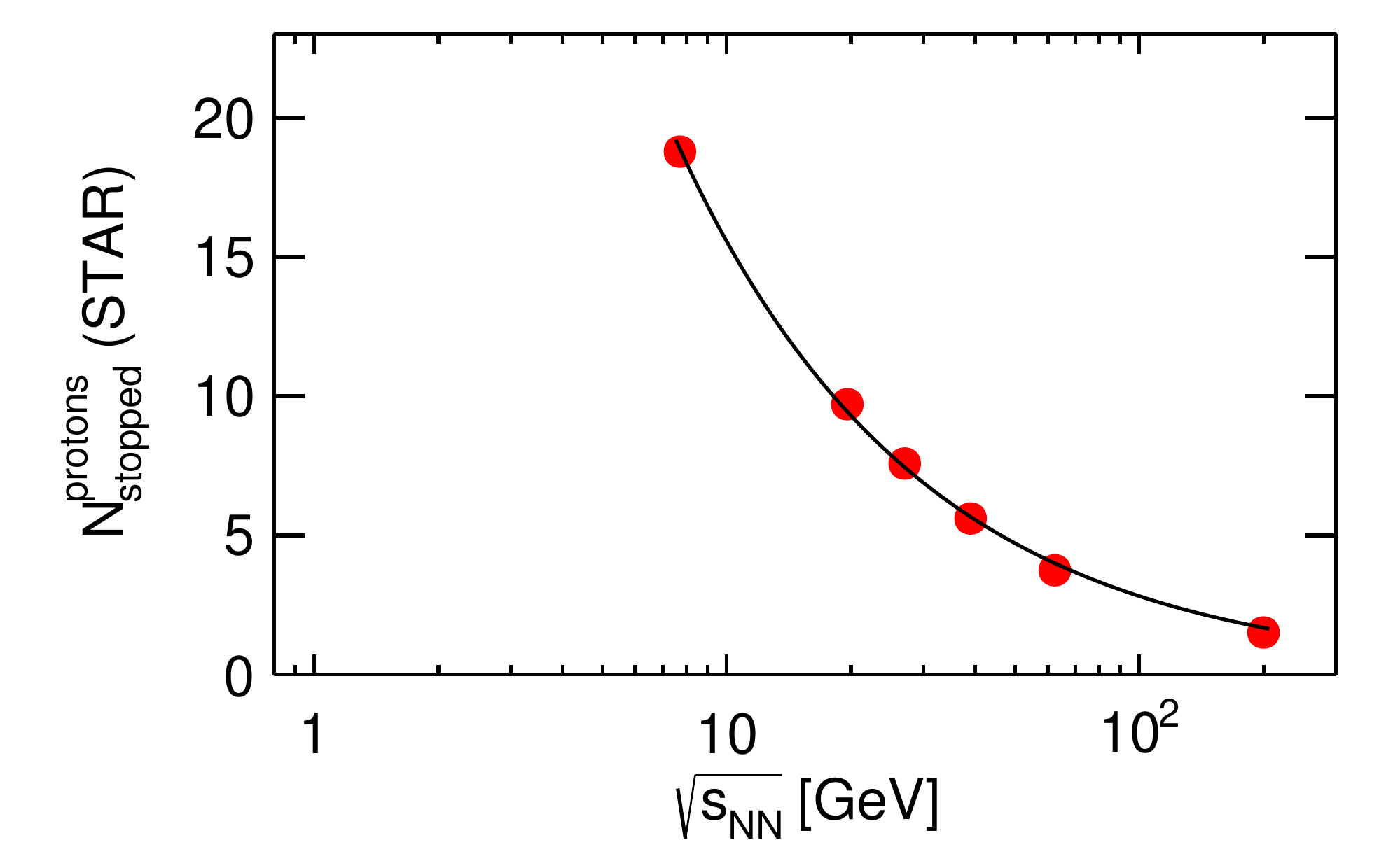}
\caption{ Mean number of protons from baryon stopping  in STAR acceptance as a function of $\sqrt{s_{\rm NN}}$, showing an exponential decrease with energy.}
\label{fig5}
\end{figure}

  FIG.~\ref{fig5} shows the energy dependent behaviour of the stopped protons, which is consistent with the fact that stopping is more at lower energies as compared to the higher energies. 

 It is interesting to note that after subtracting the stopped protons from the mean of STAR proton distributions~\cite{Adamczyk:2013dal}, the remaining produced protons are consistent with the antiprotons measured by the STAR experiment. This can be seen from TABLE.~\ref{table5}. There is a small discrepancy between these two numbers at 7.7 GeV and 11.5 GeV measurements. This may be due to the larger uncertainties in the experimental measurement of protons itself. 

 \begin{table}[H]
 \centering
   \caption{Column wise: (a) $\sqrt{s_{\rm NN}}$ (in GeV) at which the analysis is performed, (b) Mean number of protons obtained from baryon stopping in STAR acceptance [$N^{\rm protons}_{\rm stopped}(\rm STAR)$], (c) Mean number of protons measured by STAR experiment [$N^{\rm protons}_{\rm STAR}$]  \cite{stardata}, (d) Diff. =[$N^{\rm protons}_{\rm STAR}$-$N^{\rm protons}_{\rm stopped}(\rm STAR)$], and (e) Mean number of antiprotons measured by STAR experiment [$N^{\rm antiprotons}_{\rm STAR}$].}

\small
\begin{tabular}{|c|c|c|c|c|c|c|}
\hline
 (a) &         (b)        & (c)                       & (d)  & (e)     \\ \hline
 $\sqrt{s_{\rm NN}}$ & $\rm N^{\rm protons}_{\rm stopped}(\rm STAR)$  &~$\rm N^{\rm protons}_{\rm STAR}$& Diff. & $\rm N^{\rm antiprotons}_{\rm STAR}$    \\ \hline
  7.7          &17.21 $\pm$ 0.86     &18.92 $\pm$0.01       &1.71$\pm$ 0.86      &0.165          \\ \hline 
  11.5       &12.89 $\pm$ 0.86     &15.00 $\pm$0.01       &2.10$\pm$ 0.86            &0.49          \\ \hline
  19.6      &9.73  $\pm$ 0.80     &11.37$\pm$0.00         &1.63$\pm$ 0.80           &1.15           \\ \hline
  27.0     &7.61   $\pm$ 0.73     &9.39$\pm$0.00          &1.78 $\pm$ 0.73          &1.65           \\ \hline
  39.0     &5.78 $\pm$  0.65    &8.22$\pm$0.00           &2.44$\pm$  0.65             &2.38           \\ \hline
  62.4    &3.78  $\pm$ 0.54    &7.25$\pm$0.00            &3.47$\pm$ 0.54              &3.14            \\ \hline
  200    &1.54  $\pm$ 0.33  &5.664$\pm$0.00              &4.12$\pm$ 0.33            &4.11           \\ \hline
    \end{tabular}
\label{tablefin}
\label{table5}
\end{table}
\section{Experimental Implications}
\label{sec:expt}
The discussed methodology would lead to the estimation of the number of stopped protons that could improve the understanding of dynamical fluctuations. Particularly, the $\kappa\sigma^{2}$ variable of the produced protons {\it only}, will be useful to study the criticality in the QCD phase diagram. As, it is not possible to tag a proton from stopping or production in experimental data, the correction for the stopped protons to the net-proton multiplicity distribution can't be applied to the experimental measurement. The exact distribution of stopped protons on an event-by-event basis is not known and hence, it is difficult to subtract the contribution of stopped protons to the inclusive proton distribution, which needs further investigations. Therefore, it is suggested that, while quoting the fluctuation results from net-proton multiplicity distribution, a systematic uncertainty may be added by using the Monte Carlo simulations and the results of this study to quantify the effect of stopping on the fluctuations in their respective acceptance. This proposed analysis bears values in the low energy BES program of RHIC, where a search for critical point and the associated criticality becomes prudently viable.

\section{Summary} 
\label{sec:sum}   
In the present work, we use the data of net-proton rapidity distributions for the most central \auau collisions measured by different experiments. Further, we use a two source function~\cite{Ivanov:2010cu}, to analyze the net-proton distributions and the percentage of stopped protons at different center-of-mass energies. Then the percentage of stopped protons is interpolated at RHIC  BES energies.  Afterwords, using invariant transverse momentum spectrum we estimate the fraction of the number of protons contributing in STAR acceptance.  Finally, using these two numbers, we estimate the contribution of stopped protons in $|y| ~< ~0.5$ and transverse momentum between 0.4 GeV/c and 0.8 GeV/c, which is used to measure the protons and antiprotons by STAR experiment for net-proton fluctuation studies~\cite{Adamczyk:2013dal}. The critical point of QCD phase diagram is expected to show large dynamical fluctuations in the produced conserved charges. Therefore, it will be exciting to see these results after removing the contribution of stopped protons, which have significant contributions, particularly at lower collision energies. 

\section{Acknowledgement} 
Dhananjaya Thakur acknowledges the financial support from the University Grants Commission (UGC), New Delhi, Government of India. The authors would like to gratefully acknowledge discussions with Prof. B.K. Nandi, Dr. Dipak K. Mishra, and Dr. Sudipan De for careful reading of the manuscript.

\end{document}